\documentclass[american,english,aps,prb]{revtex4}
\usepackage[T1]{fontenc}
\usepackage[latin1]{inputenc}
\usepackage{amssymb}

\makeatletter

\providecommand{\LyX}{L\kern-.1667em\lower.25em\hbox{Y}\kern-.125emX\@}

\usepackage[bbgreekl]{mathbbol}

\usepackage{babel}
\makeatother
\begin{document}

\title{Homogeneous Nucleation in Inhomogeneous Media II: Nucleation in a
Shear Flow}

\author{David Reguera}

\email{davidr@chem.ucla.edu}

\affiliation{Department of Chemistry and Biochemistry, UCLA, 607 Charles E. Young
East Drive, Los Angeles, CA 90095-1569}

\author{J. M. Rub\'{i}}

\thanks{Permanent address: Departament de F\'{i}sica Fonamental, Universitat
de Barcelona, Diagonal 647, 08028 Barcelona, Spain}

\affiliation{Department of Chemistry, Norwegian University of Science and Technology,
7491 Trodheim, Norway}

\begin{abstract}
We investigate the influence of a shear flow on the process of nucleation.
Mesoscopic nonequilibrium thermodynamics is used to derive the Fokker-Planck
equation governing the evolution of the cluster size distribution
in a metastable phase subjected to a stationary flow. The presence
of the flow manifests itself in the expression for the effective diffusion
coefficient of a cluster and introduces modifications in the nucleation
rate. The implications of these results in condensation and polymer
crystallization are discussed.
\end{abstract}
\maketitle

\section{Introduction}

In Paper I \cite{Paper I} of this series we used mesoscopic nonequilibrium
thermodynamics (MNET) to analyze the effect of the inhomogeneities
of the medium on nucleation. Another interesting case in which  the
medium may exert an influence on the nucleation process arises from
the presence of flows or stresses in the system. This factor is specially
relevant in crystallization, which often involves mechanical processing
of the melt, such as extrusion, shearing or injection that may affect
the crystallization process drastically. Therefore it is desirable
to implement a model that takes these mechanical influences into account
in a correct and consistent way. Several models have been proposed
to capture some features of phase transitions under shear flow \cite{chap5:onuki}
but until now, there has been no theory providing a complete description
of crystallization under these conditions. 

The objective in this paper will be to analyze precisely the possible
effects of mechanical stress in nucleation and crystallization, focusing
on the simplest influence. To that end, we discuss the case of nucleation
in the presence of a stationary flow, which for simplicity is chosen
to be a shear flow. MNET \cite{chap2:entropic,chap5:temp} will be
used to derive the kinetic equation governing the dynamics of nucleation
in the presence of a shear flow, whose influence on vapor-phase nucleation
and polymer crystallization will be evaluated qualitatively.

It is worth pointing out that the presence of the flow destroys the
isotropy of the system, and leads to a distinction between diffusion
in different directions. Moreover, it introduces spatial inhomogeneities,
which induce spatial, and velocity and temperature fluxes. Coupling
between these currents may be very important, and consequently the
evolution of the probability density, velocity and temperature fields
will be governed by a highly coupled set of kinetic differential equations.
These equations can be derived within the general framework of MNET
\cite{chap2:entropic,chap5:temp}, following the steps developed in
paper I of this series\cite{Paper I}. For illustrative purposes,
it is best to restrict the analysis to effects originating purely
from the presence of the flow. Consequently, for the sake of simplicity,
we assume isothermal conditions. 

The paper is organized as follows. In Section II, we derive the Fokker-Planck
equation governing the evolution of the size and velocity distributions
of the emerging clusters in the presence of stationary shear flow.
In Section III, we simplify the description, focusing on the diffusion
regime. We thus obtain the kinetic equation governing nucleation in
the presence of a stationary flow. The implications of our results
for nucleation experiments and in polymer crystallization are discussed
briefly in Section IV. Finally, we summarize our main conclusions
in Section V.

\section{Nucleation in a stationary flow}

Consider a metastable phase in which the emerging clusters (liquid
droplets or crystallites) are embedded. Assume that this phase acts
as a heat bath imposing a constant temperature on the system. Following
the scheme developed in Paper I \foreignlanguage{american}{\cite{Paper I},
the goal of this section is the derivation of the kinetic equation
that describes nucleation in the presence of a shear flow. For this
purpose, we will work within the framework of MNET. To account for
spatial inhomogeneities, we perform a local description in terms of
$f(n,{\textbf {x}},{\textbf {u}},t)/N$, the probability density of
finding a cluster of $n\in (n,n+dn)$ molecules at position ${\textbf {x}}\in ({\textbf {x}},{\textbf {x}}+d{\textbf {x}})$,
with velocity ${\textbf {u}}\in ({\textbf {u}},{\textbf {u}}+d{\textbf {u}})$,
at time $t$.}

The starting point is the Gibbs equation for the entropy variation
of this system with respect to a reference state. In this situation,
the reference state is a stationary one characterized by a steady
shear flow velocity profile ${\textbf {v}}_{0}({\textbf {x}})$. 

Assuming local equilibrium, the entropy variation is given by the
Gibbs equation \textbf{}(see Appendix I) as

\begin{equation}
T\rho \delta s=\rho \delta e+p\rho \delta \rho ^{-1}-\rho \int \mu \delta c\, dnd{\textbf {u}}.\label{Gibbs centro de masa}\end{equation}
 Here $s({\textbf {x}},t)$ and $e({\textbf {x}},t)$ are the local
entropy and energy per cluster, respectively, $p({\textbf {x}},t)$
is the hydrostatic pressure, $\rho ({\textbf {x}},t)$ is the total
number of clusters per unit of volume, $\mu (n,{\textbf {x}},{\textbf {u}},t)$
is a non-equilibrium chemical potential, and $c(n,{\textbf {x}},{\textbf {u}},t)=f(n,{\textbf {x}},{\textbf {u}},t)/\rho $
is the number fraction of clusters. Eq. (\ref{Gibbs centro de masa})
remains valid within an element of mass following the center of mass
motion of the cluster `gas' \cite{chap5:kn:degroot}. The explicit
expression for the chemical potential can be obtained, as in paper
I, by comparison of the Gibbs equation (\ref{Gibbs centro de masa})
with the Gibbs' entropy postulate \cite{chap5:kn:degroot}, yielding

\begin{equation}
\mu (n,{\textbf {u}},{\textbf {x}},t)=k_{B}T\ln \frac{f}{f_{leq}}+\mu _{leq}({\textbf {x}}),\label{mu para feq shear}\end{equation}
 where $\mu _{leq}({\textbf {x}})$ denotes the local equilibrium
chemical potential, which is independent of the $n$ and ${\textbf {u}}$,
and $f_{leq}$ is the local equilibrium distribution function corresponding
to the reference state, a function given by\begin{equation}
f_{leq}(n,{\textbf {u}},{\textbf {x}})=\exp \left(\frac{\mu _{leq}-C(n,{u}\mathbf{)}}{k_{B}T}\right).\label{distribucion de equilibrio shear}\end{equation}
The quantity $C(n,{\textbf {u}})$ denotes the free energy of formation
of a state described by the internal variables $(n,{\textbf {u}},)$,
and is given by

\begin{equation}
C(n,{\textbf {u}})=\Delta G(n)+\frac{1}{2}m(n)({\textbf {u}}-{\textbf {v}}_{0})^{2},\label{barrera shear}\end{equation}
 where $\Delta G(n)$ represents again the energy of formation of
a cluster of size $n$ \emph{at rest} and the second term is its kinetic
energy with respect to the steady state velocity profile ${\textbf {v}}_{0}$. 

The next step toward obtaining the Fokker-Planck equation, is the
formulation of the conservation laws for the `gas' of clusters. In
the absence of external body forces, the distribution function obeys
the following continuity equation 

\begin{equation}
\frac{\partial f}{\partial t}=-{\textbf {u}}\cdot \nabla f=-\frac{\partial }{\partial {\textbf {u}}}\cdot {\textbf {J}}_{u}-\frac{\partial J_{n}}{\partial n},\label{ec continuidad}\end{equation}
 which introduces the phase space current ${\textbf {J}}_{u}$, arising
from the interaction of the clusters with the surrounding metastable
phase. 

In order to derive the entropy balance equation, we also need the
expression for the variation of energy with respect to its value at
the stationary state. The presence of the external flow is responsible
for the appearance in that equation of a term for `viscous heating',
that gives rise to variations in the temperature field. To maintain
isothermal conditions, and following ideas introduced previously in
the implementation of the so-called `homogeneous shear' \textbf{}\cite{chap5:hover,chap5:keizer},
we assume the existence of a local heat source capable of removing
heat generated in the process. Under this assumption, the energy of
a volume element remains constant along its trajectory so that its
balance equation can be omitted in the following analysis.

Introducing into the Gibbs equation (\ref{Gibbs centro de masa})
the equation for number fraction 

\begin{equation}
\rho \frac{dc}{dt}=-\nabla \cdot f({\textbf {u}}-{\textbf {v}}_{0})-\frac{\partial }{\partial {\textbf {u}}}\cdot {\textbf {J}}_{u}-\frac{\partial J_{n}}{\partial n},\label{ec para fraccion masica}\end{equation}
 derived in Appendix II, we obtain, after some straightforward algebraic
manipulations, the entropy balance equation 

\begin{equation}
\rho \frac{ds}{dt}=-\nabla \cdot {\textbf {J}}_{s}+\sigma ,\label{balance de entropia shear}\end{equation}
 where the entropy flux ${\textbf {J}}_{s}$ is given by 

\begin{equation}
{\textbf {J}}_{s}=-k_{B}\int f(lnf-1)({\textbf {u}}-{\textbf {v}}_{0})dnd{\textbf {u}}-\frac{1}{T}\int C(n,{\textbf {u}})f({\textbf {u}}-{\textbf {v}}_{0})^{2}dnd{\textbf {u}},\label{flujo total de entropia}\end{equation}
and the entropy production, $\sigma $, which must be positive according
to the second law of thermodynamics, is

\begin{eqnarray}
\sigma =-\frac{1}{T}\int {\textbf {J}}_{u}\cdot \frac{\partial \mu }{\partial \textbf {u}}dnd{\textbf {u}}-\frac{1}{T}\int J_{n}\frac{\partial \mu }{\partial n}\, dnd{\textbf {u}} &  & \nonumber \\
-\frac{1}{T}\int f\, ({\textbf {u}}-{\textbf {v}}_{0})\cdot \nabla \left(\frac{1}{2}m(n)({\textbf {u}}-{\textbf {v}}_{0})^{2}\right)dnd{u}\mathbf{.} &  & \label{produccion de entropia shear}
\end{eqnarray}

In order to obtain the entropy flux we have additionally employed
the identity

\begin{equation}
\int \nabla \cdot \left(({\textbf {u}}-{\textbf {v}}_{0})f\right)\ln f\, dnd{\textbf {u}}=\nabla \cdot \int ({\textbf {u}}-{\textbf {v}}_{0})f(\ln f-1)\, dnd{u}\mathbf{-}\rho \nabla \cdot {\textbf {v}}_{0}\, \mathbf{,}\label{auxiliar}\end{equation}
 and have omitted, in the entropy production, the term$-(k_{B}\rho -\frac{p}{T})\nabla \cdot {\textbf {v}}_{0}$
arising from the bulk viscosity, which, in general, is negligibly
small. 

The entropy production can be interpreted as a sum of conjugate forces-fluxes
pairs. In the present situation, the forces are $-\frac{1}{T}\frac{\partial \mu }{\partial {\mathbf{u}}}$,
$-\frac{1}{T}\frac{\partial \mu }{\partial n}$, and $-\frac{1}{T}\nabla \left(\frac{1}{2}m(n)({\textbf {u}}-{\textbf {v}}_{0})^{2}\right)$
while their conjugated fluxes are the velocity flux ${\textbf {J}}_{u}$,
the flux in size space $J_{n}$, and the relative spatial current
${\textbf {J}}_{x}\equiv f({\textbf {u}}-{\textbf {v}}_{0})$, respectively. 

The next step is the formulation of coupled linear phenomenological
equations relating these fluxes and forces. The situation under consideration
presents some subtleties. Spatial and velocity currents have the same
tensorial rank (they are both vectors) which implies that, in general,
they must be coupled. Moreover, the presence of shear flow destroys
the isotropy of the system by introducing a special direction. Consequently,
in general, phenomenological coefficients are expected to be tensors.
In addition, locality in the internal space will again be assumed
. Taking these considerations into account, the expressions for the
currents are\begin{equation}
{\textbf {J}}_{x}=+\frac{m}{T}\, {\mathcal{L}}_{xx}\, \cdot \nabla {\textbf {v}}_{0}\cdot ({\textbf {u}}-{\textbf {v}}_{0})+\frac{1}{T}\, {\mathcal{L}}_{ux}\cdot \frac{\partial \mu }{\partial \textbf {u}},\label{ley fenome x shear}\end{equation}

\begin{equation}
{\textbf {J}}_{u}=-\frac{1}{T}\, {\mathcal{L}}_{uu}\, \cdot \frac{\partial \mu }{\partial \textbf {u}}+\frac{m}{T}{\mathcal{L}}_{ux}\cdot \nabla {\textbf {v}}_{0}\cdot ({\textbf {u}}-{\textbf {v}}_{0}),\label{ley fenome u shear}\end{equation}

\begin{equation}
J_{n}=-\frac{1}{T}L_{nn}\, \frac{\partial \mu }{\partial n},\label{ley fenome gamma shear}\end{equation}
where we have used the Onsager reciprocal relation \begin{eqnarray}
{\mathcal{L}}_{ux}=-{\mathcal{L}}_{xu}\, , &  & \label{Onsager shear}
\end{eqnarray}
as well as the result

\begin{equation}
\nabla \left(\frac{1}{2}m({\textbf {u}}-{\textbf {v}}_{0})^{2}\right)=-m\nabla {\textbf {v}}_{0}\cdot ({\textbf {u}}-{\textbf {v}}_{0}).\label{calculillo}\end{equation}

It is useful to redefine the phenomenological coefficients in a more
convenient way. By identifying $D_{n}\equiv \frac{k_{B}L_{nn}}{f}$
as the diffusion coefficient in $n$-space, which in turn can be identified
with $k^{+}(n)$, the rate of attachment of monomers to a cluster,
while introducing the \emph{friction tensors} $\overrightarrow{\overrightarrow{\alpha }}=\frac{m{\mathcal{L}}_{uu}}{fT}$
and $\overrightarrow{\overrightarrow{\zeta }}=\frac{m{\mathcal{L}}_{ux}}{fT}$
as well as the explicit form of the chemical potential, equation (\ref{mu para feq shear}),
the currents ${\textbf {J}}_{u}$ and $J_{n}$ can be written as

\begin{equation}
{\textbf {J}}_{u}=-f\left[\overrightarrow{\overrightarrow{\alpha }}+\overrightarrow{\overrightarrow{\zeta }}\cdot \nabla {\textbf {v}}_{0}\right]\cdot ({\textbf {u}}-{\textbf {v}}_{0})-\overrightarrow{\overrightarrow{\alpha }}\frac{k_{B}T}{m}\cdot \frac{\partial f}{\partial {u}},\label{flujo de u final shear}\end{equation}

\begin{equation}
J_{n}=-D_{n}\left(\frac{\partial f}{\partial n}+\frac{1}{k_{B}T}\frac{\partial C}{\partial n}f\right).\label{flujo de g final shear}\end{equation}

Introduction of these expressions for the currents into the continuity
equation yields 

\begin{eqnarray}
\frac{\partial f}{\partial t}=-{\textbf {u}}\cdot \nabla f+\frac{\partial }{\partial n}\left[D_{n}\left(\frac{\partial f}{\partial n}+\frac{1}{k_{B}T}\frac{\partial C}{\partial n}f\right)\right]+ &  & \nonumber \\
+{\frac{\partial }{\partial {\textbf {u}}}}\cdot \left(\left[\overrightarrow{\overrightarrow{\alpha }}+\overrightarrow{\overrightarrow{\zeta }}\cdot \nabla {\textbf {v}}_{0}\right]\cdot f({\textbf {u}}-{\textbf {v}}_{0})+\overrightarrow{\overrightarrow{\alpha }}\frac{k_{B}T}{m}\cdot \frac{\partial f}{\partial {u}}\right), &  & \label{FP shear}
\end{eqnarray}
which is the Fokker-Planck equation describing the evolution of the
inhomogeneous density distribution of clusters in the presence of
steady flow.

\section{Balance Equations in the Diffusion Regime}

The Fokker-Planck equation, Eq. (\ref{FP shear}), we have derived
retains information concerning the evolution of the cluster velocity
distribution. Although the movement of clusters may play a significant
role in the nucleation process, as discussed in Ref. \cite{chap4:noneq-trans-rot},
velocity dependences can hardly be measured because velocity distribution
relaxes to equilibrium so rapidly. Therefore, as discussed in Paper
I, a simplified description of the process can be provided, by coarse-graining
the dependence on the velocity variables. The relevant quantities
are then the different moments of the distribution function, namely
the density of clusters

\selectlanguage{american}
\begin{equation}
f_{c}(n,{\textbf {x}},t)=\int f(n,{\textbf {x}},{\textbf {u}},t)d{\textbf {u}}\: ,\label{densidad browniana}\end{equation}
 the coarse-grained velocity of the clusters

\begin{equation}
{\textbf {v}}_{c}(n,{\textbf {x}},t)=f_{c}^{-1}\int {\textbf {u}}fd{\textbf {u}}\: ,\label{densidad de momento}\end{equation}
and the second moment  

\begin{equation}
{\mathcal{P}}=\int f({\textbf {u}}-{\textbf {v}}_{c})({\textbf {u}}-{\textbf {v}}_{c})d{\textbf {u}},\label{definicion del tensor de presiones}\end{equation}
which is related to the kinetic definition of the pressure tensor
\cite{chap5:kn:degroot,chap5:chapman,chap5:hirschfelder}. 

\selectlanguage{english}
Proceeding along the lines of Section III of Paper I, integration
of Eq. (\ref{FP shear}) over velocities leads to the corresponding
balance equations, namely

\begin{equation}
\frac{\partial f_{c}}{\partial t}=-\nabla \cdot f_{c}{\textbf {v}}_{c}-\frac{\partial }{\partial n}\int J_{n}d{\textbf {u}}\: ,\label{continuidad shear}\end{equation}
 for the density of clusters in n-space, \begin{equation}
f_{c}\frac{d{\textbf {v}}_{c}}{dt}=-\nabla \cdot {\mathcal{P}}-{\mathcal{B}}\cdot f_{c}\left({\textbf {v}}_{c}-{\textbf {v}}_{0}\right)-\int ({\textbf {u}}-{\textbf {v}}_{c})\frac{\partial }{\partial n}J_{n}d{\textbf {u}},\label{balance de v shear}\end{equation}
for the momentum, and 

\begin{eqnarray}
\frac{d}{dt}{\mathcal{P}}=-\nabla \cdot {\mathcal{Q}}-2({\mathcal{P}}\cdot \nabla {\textbf {v}}_{c})^{s}-{\mathcal{P}}\nabla \cdot {\textbf {v}}_{c}-2\left({\mathcal{B}}\cdot {\mathcal{P}}\right)^{s} & -\frac{2k_{B}T}{m}f_{c}\left(\overrightarrow{\overrightarrow{\alpha }}\right)^{s} & \nonumber \\
-\int ({\textbf {u}}-{\textbf {v}}_{c})({\textbf {u}}-{\textbf {v}}_{c})\frac{\partial }{\partial n}J_{n}d{\textbf {u}}, &  & \label{balance de p shear}
\end{eqnarray}
 for the pressure tensor, where ${\mathcal{Q}}$ is, as in paper I
\cite{Paper I}, the kinetic part of the heat flux and ${\mathcal{B}}\equiv \overrightarrow{\overrightarrow{\alpha }}+\overrightarrow{\overrightarrow{\zeta }}\cdot \nabla {\textbf {v}}_{0}$.

A remarkable feature of these equations is the fact that the balance
equation for the pressure tensor, Eq. (\ref{balance de p shear}),
is similar to the one obtained for the thirteen moment approximation
to the Boltzmann equation \cite{chap5:hirschfelder}. It is thus worth
pointing out that the mesoscopic nonequilibrium thermodynamic treatment
of the problem has been able to recover results similar to those of
kinetic theory. In particular, inertial regimes and relaxation equations
for the main quantities can be obtained within this framework. This
result reinforces the fact that application of the well-established
postulates of non-equilibrium thermodynamics as indicated in Ref.\cite{chap5:kn:degroot},
suffices to provide a general scheme under which non-equilibrium processes
of both macroscopic and mesoscopic nature can be treated.

The second term on the right hand side of the momentum balance equation
can be identified with the frictional force per unit mass exerted
on the cluster suspended in the host fluid. ${\mathcal{B}}$ plays
the role of the friction tensor. The modulus of this tensor again
determines the characteristic time scale for the relaxation of the
velocities. In general, such a friction tensor can be calculated from
hydrodynamics; however, for the sake of simplicity, we approximate
this tensor by ${\mathcal{B}}\simeq \beta {\mathbb{1}},$ the Stokes
diagonal friction. This identification leads to $\vec{\vec{\alpha }}=\beta {\mathbb{1}}-\overrightarrow{\overrightarrow{\zeta }}\cdot \nabla {\textbf {v}}_{0}$. 

The inverse of the friction coefficient $\beta ^{-1}$ again shows
the separation of the dynamics of the system into a short-time inertial
regime and the longer time diffusional regime. For times $t>\beta ^{-1}$
the system enters the diffusional regime. In this situation, inertial
effects manifested through the presence of the time derivative $\frac{d{\textbf {v}}_{c}}{dt}$
can be neglected. Moreover, the contribution arising from the current
$J_{n}$ in the velocity relaxation can be neglected since the rate
of relaxation of velocities is usually faster than that of cluster
sizes determined by $D_{n}$.

The results derived thus far are valid for any stationary flow profile
${\textbf {v}}_{0}$. However, to proceed further, and for the sake
of simplicity, we will focus on the particular interesting case of
a shear flow.

Proceeding along the lines indicated in Paper I and in Ref. \cite{chap5:ivan},
we find that, in the diffusion regime, the expression for the pressure
tensor with a shear flow present, is

\begin{equation}
{\mathcal{P}}=\frac{k_{B}T}{m}f_{c}\, \left[{\mathbb{1}}-\left(\beta ^{-1}\left({\mathbb{1}}+\overrightarrow{\overrightarrow{\zeta }}\right)\cdot \nabla {\textbf {v}}_{0}\right)^{s}\right].\label{tensor P en shear (reg. dif)}\end{equation}
 From this equation, we conclude that Brownian motion of the particles
contributes to the total pressure tensor of the suspension in two
ways. The first contribution is the well-known scalar kinetic pressure
given by

\begin{equation}
p=\frac{k_{B}T}{m}f_{c},\label{ec-estado}\end{equation}
 which is the equation of state for an ideal `gas' of clusters. The
second contribution comes from the irreversible part $\Pi ^{0}$ of
the pressure tensor, a part that can be written in the form

\begin{equation}
{\Pi ^{0}}=-D_{0}f_{c}\left[({\mathbb{1}}+\overrightarrow{\overrightarrow{\zeta }})\cdot \nabla {\textbf {v}}_{0}\right]^{0},\label{pi griega brownies}\end{equation}
 where $D_{0}=\frac{k_{B}T}{m\beta }$ is the diffusion coefficient
of the cluster for the liquid at rest, and the superscript $0$ indicates
a symmetric traceless tensor. This last equation defines the Brownian
viscosity tensor

\begin{equation}
\vec{\vec{\eta }}_{B}=D_{0}f_{c}({\mathbb{1}}+\overrightarrow{\overrightarrow{\zeta }}),\label{viscosidad}\end{equation}
 which contains the {}``Brownian viscosity'' $D_{0}f_{c}$ \cite{chap5:freed},
and the contribution due to coupling with the non-equilibrium bath,
proportional to $\overrightarrow{\overrightarrow{\zeta }}$. 

Inserting the expression for the pressure tensor in the velocity balance
equation (\ref{balance de v shear}), and taking the diffusion limit
for the cluster current we obtain

\begin{equation}
{\textbf {J}}_{D}=f_{c}{\textbf {v}}_{c}=-{\mathcal{D}}\cdot \nabla f_{c}+f_{c}{\textbf {v}}_{0}\label{JD shear dif}\end{equation}
where \begin{equation}
{\mathcal{D}}=D_{0}\left[{\mathbb{1}}-\left(\beta ^{-1}\left({\mathbb{1}}+\overrightarrow{\overrightarrow{\zeta }}\right)\cdot \nabla {\textbf {v}}_{0}\right)^{0}\right]\label{coef dif en el shear}\end{equation}
 can be identified with the \emph{spatial} diffusion coefficient,
which in this situation possesses tensorial character. Using Eqs.
(\ref{viscosidad}) and (\ref{ec-estado}), ${\mathcal{D}}$ can be
rewritten as

\begin{equation}
{\mathcal{D}}=D_{0}\left[{\mathbb{1}}-\left(\frac{\overrightarrow{\overrightarrow{\eta }}_{B}}{p}\cdot \nabla {\textbf {v}}_{0}\right)^{0}\right].\label{coef dif shear con viscosidad}\end{equation}

Inserting this expression into the equation for the density, Eq. (\ref{continuidad shear}),
one finally obtains

\begin{equation}
\frac{\partial f_{c}}{\partial t}=-\nabla \cdot \left(f_{c}{\textbf {v}}_{0}\right)+\nabla \cdot \left({\mathcal{D}}\cdot \nabla f_{c}\right)+\frac{\partial }{\partial n}\left(D_{n}\frac{\partial f_{c}}{\partial n}+\frac{D_{n}}{k_{B}T}\frac{\partial \widetilde{\Delta G}}{\partial n}f_{c}\right),\label{densidad browniana en el shear (dif)}\end{equation}
which is the kinetic equation for nucleation in presence of a shear
flow in the \emph{diffusion regime}, and where, as discussed in Paper
I, $\widetilde{\Delta G}(n)$ is the effective nucleation barrier
in the diffusion regime given by 

\selectlanguage{american}
\begin{equation}
\widetilde{\Delta G}(n)=\Delta G(n)+\frac{3}{2}k_{B}T\ln n,\label{barrera corregida}\end{equation}

\selectlanguage{english}
\section{Influence of Shear Flow in Nucleation and Crystallization}

In this section, we analyze the conditions under which mechanical
stress that cause a shear flow in the system, may influence the nucleation
process. Obviously, a shear rate modifies transport and results in
alterations in density, pressure and temperature profiles. To isolate
the influence of mechanical stress, heat and mass transfer will be
neglected. The focus will be on the direct effect of flow on nucleation
rate, in both condensation and crystallization, with special emphasis
on polymer crystallization.

\subsection{Influence on Condensation }

As shown in the previous section, an important influence of flow on
nucleation is exercised through its effect on the diffusion coefficient
as exhibited by Eq. (\ref{coef dif shear con viscosidad}). Typically,
nucleation rates $J$ are given by \cite{chap1:kn:talan,chap1:kn:Kelton}

\selectlanguage{american}
\begin{equation}
J\sim k^{+}(n^{*})e^{-\frac{\widetilde{\Delta G}(n*)}{k_{B}T}},\label{j clas}\end{equation}
\foreignlanguage{english}{where $k^{+}(n)$ is the rate of attachment
of molecules to a $n-$cluster, and $\widetilde{\Delta G}(n*)$ is
the height of the nucleation barrier (i.e. the free energy of formation
of the critical cluster of size $n^{*}$). The significance of the
modification of the diffusion coefficient lies in the fact that the
rate of attachment of molecules to the nucleus, $k^{+}$, and thus
the pre-exponential factor in the nucleation rate, Eq. (\ref{j clas}),
is roughly proportional to the the diffusivity of molecules \cite{chap5:kn:greer}.
The fact that the diffusion coefficient could be altered by the flow
implies that the nucleation rate is modified accordingly. The modification,
prescribed by Eq. (\ref{coef dif shear con viscosidad}), depends
on the shear rate, the pressure and the viscosity, and its magnitude
is expected to be important when }

\selectlanguage{english}
\begin{equation}
\frac{\eta }{p}\left|\nabla {\textbf {v}}_{0}\right|\geq 1,\label{importancia del shear}\end{equation}
i.e., for high enough values of viscosity, and shear rate, and for
low pressures. 

To be more precise, let us estimate the magnitude of the correction
for a typical experiment in a laminar flow diffusion cloud chamber
\cite{chap5:pentanol laminar}. Representative values of the shear
rates, the pressure and the viscosity in these experiments are $\nabla v_{0}\sim 10^{3}\; s^{-1}$,
$p\sim 10^{4}\: dyn/cm^{2}$, and $\eta \sim 10^{-4}\: poise$. These
numbers yield an approximate value of $10^{-5}$ for the order of
magnitude of the correction introduced by the shear. Consequently,
in this range of values, neither diffusion nor the nucleation rate
are significantly altered by shear.

\subsection{Influence in Polymer Crystallization}

Whereas in condensation under normal conditions, the modification
of the diffusion coefficient due to the shear flow is in general negligible,
this is not the case in crystallization and specially in polymer crystallization.
The presence of flow affects crystallization in different ways. On
the one hand, nucleation rates can be directly modified by flow. Typical
values of shear and injection rates, and of the mechanical stress
applied to a polymer melt during crystallization are usually rather
large. In addition, the viscosity of the supercooled melt can also
be very large, specially near the glass transition. In this situation,
the adjustment of the diffusion coefficient, given by Eq. (\ref{coef dif shear con viscosidad}),
can be very significant and thus modify the nucleation rate dramatically. 

On the other hand, the symmetry breaking inherent in the flow causes
the diffusion coefficient to no longer be a scalar and to develop
a \emph{tensorial} nature. The rate of the process then depends on
direction so that nucleation, and specially the crystal growth, are
no longer isotropic. 

Finally, another important factor to be considered is that flow also
induces stress on the surface of the cluster that may fragment droplets
that become large enough \cite{chap5:min,chap5:harrowell}. This effect
can be easily incorporated into our description by including in the
free energy of cluster formation the additional energetic cost associated
with the shear stress.

\section{Conclusions}

The main objective of this paper has been the analysis of the kinetics
of nucleation and crystallization when the metastable phase is subjected
to a flow resulting from application of mechanical stress. Using MNET,
the kinetic equation that describes the evolution of the cluster size
distribution in a stationary flow was derived. For simplicity, our
results were speciallized to the case of a shear flow, and its influence
on condensation and crystallization was evaluated.

The main effects that a shear flow exerts on nucleation can be summarized
as follows. First, such flow alters transport and consequently the
evolution of the cluster distribution function, that among other things,
determines nucleation and growth rates. Second, flow changes the cluster
spatial diffusion coefficient in accordance with equation (\ref{coef dif en el shear}).
Typical values of the parameters controlling this adjustment indicate
that the effect is not very important in condensation. However, high
viscosity and other peculiarities of polymer crystallization suggest
that a shear may promote drastic changes in the process. The formalism
developed in this paper can help to explain experimental results of
polymer crystallization under shear \cite{chap5:eder}. However, quantitative
prediction is limited, for the moment, by the unavailability of the
magnitudes of the physical parameters required in calculation. Measurement
of these parameters has to be performed before a complete characterization
of the effect of flow, using our theory, on the kinetics of polymer
crystallization can be explicitly evaluated.

As indicated, the main object of this paper has been the analysis
of the effects that flow by itself may induce on the nucleation process.
For this purpose, isothermal conditions were assumed. A more complete
and more realistic treatment that models mechanical processing of
the melt may be carried out by incorporating thermal effects and their
coupling to the shear flow, in the manner discussed in Paper I.

\begin{acknowledgments}
The authors would like to acknowledge Prof. H. Reiss for his comments
and his careful revision of the manuscript, and I. Santamaria-Holek
for valuable discussions. This work has been partially supported by
the National Science Foundation under NSF grant No. CHE-0076384\foreignlanguage{american}{,
and by} DGICYT of the Spanish Government under grant PB2002-01267.
\end{acknowledgments}
\newpage

\section*{Appendix A}

\subsection*{Local Thermodynamic Relations}

The Gibbs equation for a multicomponent thermodynamic system is\cite{chap5:kn:degroot}

\begin{equation}
T\delta S=\delta E+p\delta V-\sum \mu \delta N.\label{ap:gibbs}\end{equation}
In the continuous limit this expression becomes

\begin{equation}
T\delta S=\delta E+p\delta V-\int \mu \delta N(n,{\textbf {u}})\, dnd{\textbf {u}},\label{ap: gibbs cont}\end{equation}
 where different values of $n$ and ${\textbf {u}}$ denote the equivalent
of different {}``species'', and $N(n,{\textbf {u}})dnd{\textbf {u}}$
represents the number of particles of a given {}``species''. 

Now construct the local version of Eq. (\ref{ap: gibbs cont}). For
this purpose, we will describe the system in terms of $f(n,{\textbf {x}},{\textbf {u}},t)$,
the number density of clusters of size $n\in (n,n+dn)$ at ${\textbf {x}}\in ({\textbf {x}},{\textbf {x}}+d{\textbf {x}})$,
having velocity ${\textbf {u}}\in ({\textbf {u}},{\textbf {u}}+d{\textbf {u}})$,
at time $t$. We define the specific thermodynamic quantities $s$,
$e$, $v$, and $c$ as follows

\begin{equation}
S=\int s\rho d{\textbf {x}}\label{s local}\end{equation}
\begin{equation}
E=\int e\rho d{\textbf {x}}\label{e local}\end{equation}
\begin{equation}
V=\int 1d{\textbf {x}}=\int v\rho d{\textbf {x}}\label{v local}\end{equation}
\begin{equation}
N(n,{\textbf {u}})=\int fd{\textbf {x}}=\int c\rho d{\textbf {x}}\label{n local}\end{equation}
 where $\rho ({\textbf {x}},t)=\int f\, dnd{\textbf {u}}$ is the
number density of clusters per unit volume at ${\textbf {x}}$, while
$v=\rho ^{-1},$ and $c=\frac{f}{\rho }$. 

Using these quantities, the local expression of the Gibbs equation
becomes\begin{equation}
T\delta (\rho s)=\delta (\rho e)+p\delta (1)-\int \mu \delta f\, dnd{\textbf {u}},\label{ap: Gibbs local T}\end{equation}
which is the form used in Paper I.

Alternatively, this equation can be written as

\begin{equation}
T\rho \delta s-\rho \delta e-p\rho \delta \rho ^{-1}+\rho \int \mu \delta c\, dnd{\textbf {u}}+\left(Ts-e-p\rho ^{-1}+\int \mu c\, dnd{\textbf {u}}\right)\delta \rho =0,\label{ap: intermedio}\end{equation}
which, using the definition of the Gibbs energy to cancel the last
term, reduces to\begin{equation}
T\rho \delta s=\rho \delta e+p\rho \delta \rho ^{-1}-\rho \int \mu \delta c\, dnd{\textbf {u}}.\label{ap: Gibbs local v}\end{equation}
which is the expression employed in the present paper.

\section*{Appendix B}

\subsection*{Balance Equation for the Number Fraction}

Integration of Eq. (\ref{ec continuidad}) with respect to the cluster
velocity ${\textbf {u}}$ and size $n$, leads to the macroscopic
equation of continuity, which can be written in the form 

\begin{equation}
\frac{d\rho }{dt}=-\rho \nabla \cdot {\textbf {v}},\label{balance de masa1}\end{equation}
 where $\rho ({\textbf {x}},t)$ is the density of the cluster `gas',
given by

\begin{equation}
\rho ({\textbf {x}},t)=\int f\, dnd{\textbf {u}}.\label{definicion densidad}\end{equation}
 In Eq. (\ref{balance de masa1}),${\textbf {v}}({\textbf {x}},t)$
\textbf{}is the average velocity of the Brownian particles defined
by the expression 

\begin{equation}
\rho {\textbf {v}}({\textbf {x}},t)=\int {\textbf {u}}f\, dnd{\textbf {u}},\label{definicion flujo difusion}\end{equation}
and where the {}``total derivative'' is \textbf{}\begin{equation}
\frac{d}{dt}\equiv \frac{\partial }{\partial t}+{\textbf {v}}\cdot \nabla .\label{derivada total}\end{equation}

Using the continuity equation, Eq. (\ref{ec continuidad}), and Eq.
(\ref{balance de masa1}), one obtains the equation for the number
fraction\begin{equation}
\rho \frac{dc}{dt}=\frac{df}{dt}-\frac{f}{\rho }\frac{d\rho }{dt}=-\nabla \cdot f({\textbf {u}}-{\textbf {v}})-\frac{\partial }{\partial {\textbf {u}}}\cdot {\textbf {J}}_{u}-\frac{\partial J_{n}}{\partial n}.\label{balance number fraction}\end{equation}
Under the assumption of local equilibrium , the average velocity ${\textbf {v}}$
coincides with the velocity of the reference state, i.e. ${\textbf {v}={\textbf {v}_{0}}}$,
thus leading to Eq. (\ref{ec para fraccion masica}). 

\newpage

\section*{List of References}

\end{document}